\documentclass[preprint,5p,times,twocolumn]{elsarticle}

\usepackage{graphics}
\usepackage{subfigure}
\usepackage{amssymb}
\usepackage{multicol}
\usepackage{multirow}
\usepackage{tabularx}
\usepackage{amsmath}
\usepackage{cuted}
\usepackage{lipsum}
\usepackage[dvipsnames]{xcolor}
\usepackage{soul}
\usepackage{hyperref}
\begin{document}
\begin{frontmatter}
\title{Phase transition in the majority rule model with the nonconformist agents}
\author[brin,pfisika]{Roni Muslim\corref{cp}}
\ead{roni.muslim@brin.go.id}
\cortext[cp]{Corresponding author}

\author[brin]{Sasfan A. Wella}
\ead{sasfan.arman.wella@brin.go.id}
\author[brin,3]{Ahmad R. T. Nugraha}
\ead{ahmad.ridwan.tresna.nugraha@brin.go.id}
\affiliation[brin]{
    organization={Research Center for Quantum Physics,
    National Research and Innovation Agency (BRIN)},
    city={South Tangerang},
    postcode={15314},
    country={Indonesia}
}

\affiliation[pfisika]{
    organization={Pesantren Fisika,
    Hikmah Teknosains Foundation},
    city={Yogyakarta},
    postcode={55281},
    country={Indonesia}
}
\affiliation[3]{organization={Research Collaboration Center for Quantum Technology 2.0},
    city={Bandung},
    postcode={40132}, 
    country={Indonesia}}
\begin{abstract}

Independence and anticonformity are two types of social behaviors known in social psychology literature and the most studied parameters in the opinion dynamics model. These parameters are responsible for continuous (second-order) and discontinuous (first-order) phase transition phenomena. Here, we investigate the majority rule model in which the agents adopt independence and anticonformity behaviors. We define the model on several types of graphs: complete graph, two-dimensional (2D) square lattice, and one-dimensional (1D) chain. By defining $p$ as a probability of independence (or anticonformity), we observe the model on the complete graph undergoes a continuous phase transition  where the critical points are $p_c \approx 0.334$ ($p_c\approx 0.667$) for the model with independent (anticonformist) agents. On the 2D square lattice, the model also undergoes a continuous phase transition with critical points at $p_c \approx 0.0608$ ($p_c \approx 0.4035$) for the model with independent (anticonformist) agents. On the 1D chain, there is no phase transition either with independence or anticonformity. Furthermore, with the aid of finite-size scaling analysis, we obtain the same sets of critical exponents for both models involving independent and anticonformist agents on the complete graph. Therefore they are identical to the mean-field Ising model. However, in the case of the 2D square lattice, the models with independent and anticonformist agents have different sets of critical exponents and are not identical to the 2D Ising model. Our work implies that the existence of independence behavior in a society makes it more challenging to achieve consensus compared to the same society with anticonformists.
\end{abstract}

\end{frontmatter}
\section{Introduction}

Beyond natural phenomena, many physicists have been attracted to understanding socio-political phenomena based on the concept and rules of statistical physics~\cite{galam1982sociophysics, castellano2009statistical, serge2016sociophysics}. Sociophysicists usually explain individual interactions in the real-world social system by the so-called opinion dynamics model. This model describes the interactions among the interconnected agents in the network as spin interactions in the Ising model~\cite{castellano2009statistical, noorazar2020recent}. Well-known opinion dynamics models (both continuous and discontinuous models) have been proposed thus far are the DeGroot model~\cite{degroot1974reaching}, the Friedkin–Johnsen model~\cite{friedkin1990social,johnsen1999social}, the Deffuant–Weisbuch model~\cite{deffuant2001mixing}, the Hegselmann–Krause model~\cite{hegselmann2002opinion}, the majority rule model~\cite{mobilia2003majority, galam2002minority, krapivsky2003dynamics}, the Sznajd model~\cite{sznajd2000opinion}, 
the Galam model~\cite{galam2008sociophysics}, voter or $q$-voter model~\cite{liggett1985interacting, castellano2009nonlinear},
and the Biswas-Sen model~\cite{biswas2009model}. These models are developed based on the concept of simple interactions in the Ising model, in which a microscopic setting can give rise to the phase transition of macroscopic properties. A change in physical properties characterizes this phenomenon due to small changes in "noise" parameters, for example, the temperature in the Ising model. Interestingly, it turns out that the order-disorder phase transition appears in the opinion dynamics model with the noise parameters. In this case, social behaviors such as anticonformity and independence represented by a probability $p$ act as the noise parameters which make the system undergoes an order-disorder phase transition.
Other features of statistical physics, such as scaling and universality, characterized by the universality of the critical exponents \cite{cardy1996scaling}, also emerge in the opinion dynamics models, which intrigued physicists to develop them further.

The standard approach to any opinion dynamics model is that each agent in the population obeys the prevailing norms. This kind of agent behavior refers to agents' conformity and anticonformity behaviors in the social psychology contexts. The difference between the two is that the conformist (anticonformist) will follow (contradict) the persuasion of the other agents~\cite{willis1963two, willis1965conformity}. The conformist turn the system into a homogeneous state, with all agents having the same opinion, while the anticonformist make the state of the system inhomogeneous. In statistical physics, the conformity behavior is similar to ferromagnetism, in which spins tend to become parallel due to their stable states. On the other hand, the anticonformity behavior is similar to antiferomagnetism. Anticonformity and independence behaviors are part of nonconformity behavior \cite{nail2000proposal, macdonald2004expanding}, in which the independence behavior cannot be influenced by others acting independently. The nonconformity behavior is responsible for continuous and discontinuous phase transitions of the opinion dynamics model~\cite{nyczka2013anticonformity, sznajd2011phase, crokidakis2015inflexibility, chmiel2015phase, calvelli2019phase,  abramiuk2019independence, muslim2020phase,  nowak2021discontinuous, civitarese2021external, muslim2021phase, muslim2022opinion}.

Topology is one of the essential things in terms of the occurrence of phase transition. For example, it is well-known that one cannot (can) observe phase transitions in the 1D (2D) Ising model. In sociophysics, topology describes the interaction of agents also how and with whom agents can interact in the population. This topology can refer to networks or graphs such as simple 1D chains and 2D square lattices. In a particular network, agents can interact with their neighbors, similar to the Ising spin model defined on the one and 2D regular lattice \cite{sznajd2004dynamical}. Furthermore, in the study of the disorder-disorder phase transition in the opinion dynamics model, the opinion dynamics model defined in the 2D model is found to be of the same universality class as the 2D Ising model~\cite{crokidakis2015inflexibility, calvelli2019phase, stanley1971phase, mukherjee2016disorder}.  Recent advances in information technology have also allowed one individual to connect with others, forming a complete social connection in a large-scale system \cite{trestian2017towards}. The fully connected network is a representative of a complete graph although relationships between individuals or groups can be influenced by various aspects such as social behavior, culture, and beliefs that are not as simple as imagined \cite{amblard2004role, sun2015reduced}. However, a closed community with a large member on social media can be an example of this complete connection.

In this work, we study the order-disorder phase transition of the majority rule model defined on a complete graph, 1D chain, and 2D square lattice. We consider a small group consisting of four spins or agents, and in the 2D lattice, the four agents in a square formation interact with each other based on the majority rule model. However, the number and the formation of the neighboring agent are different from the previous works \cite{crokidakis2015inflexibility, calvelli2019phase, mukherjee2016disorder}. We also estimate the critical exponents using the finite-size scaling relation to define the universality class of the model. Our results show that the model undergoes a continuous phase transition on the complete graph and 2D square lattice. No phase transition is observed for the model on the 1D chain. Moreover, the models with independent and anticonformist agents on the complete graph are identical and have the same universality class as the mean-field Ising model. Both models with independence agent and anticonformity agent on the 2D square lattice are not identical even though they are defined on the same graph. They are also not identical as two dimensional Ising model, 2D majority rule model with three agents interaction \cite{crokidakis2015inflexibility}, 2D Sznajd model \cite{calvelli2019phase}, and 2D continuous opinion dynamics model \cite{mukherjee2016disorder}.  We will further show that the probability of the system reaching a disordered state is higher in models involving independent agents than those with anticonformist agents. In other words, the value of the critical probability $p_c$ causing the system undergoes order-disorder phase transition is smaller than the critical probability of anticonformist $p_c$. These results are consistent with our previous work, although with different scenarios of agent interaction \cite{muslim2022opinion}.

\section{Model and methods}
\label{sec.2}

\begin{figure}[bt]
    \centering
    \includegraphics[width = \linewidth]{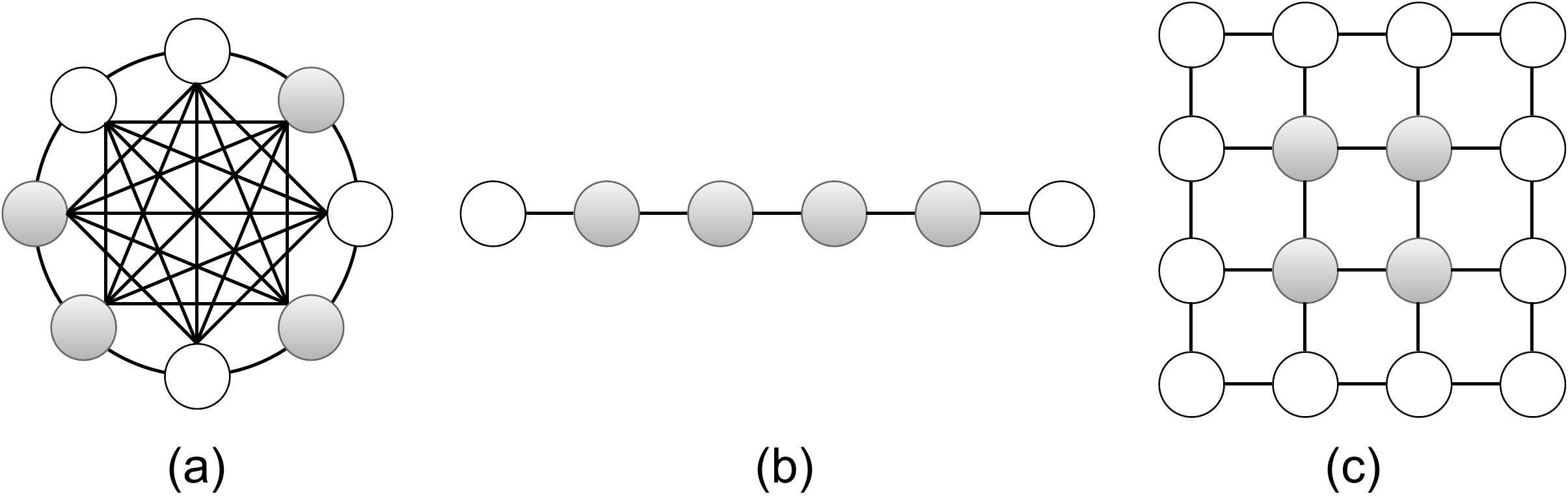}
    \caption{(a) Complete graph with eight nodes (spins or agents), (b) 1D chain, and (c) 2D square lattice. Each node embedded by Ising spin $S=\pm1$ representing an agent has two different opinions. Four agents (dark circles) are chosen randomly on the graphs and interact with each other based on the majority rule model.}
    \label{fig1}
\end{figure}

We use three simple graphs to mimic the social networks, i.e.,
complete graph, 1D chain, and 2D square lattice (see Fig.~\ref{fig1}).  Those graphs are typically used for studying phase transitions in statistical physics.  Nodes in the graph represent the agents' opinion, and the link of the graphs can be considered as a social connection, e.g. friendship.  We here consider a small group consisting of four agents, where each agent has only two possible opinions represented by Ising spin $S =\pm 1$.

The microscopic interaction of the model with independent and anticonformist agents on all three graphs can be described as follows:
\begin{enumerate}
    \item 
     Four neighboring agents in the population are chosen randomly. In the 2D square lattice, the neighboring agents are in the square formation $(S_{i,j}, S_{i+1,j}, S_{i+1,j+1}, S_{i,j+1})$.
    \item For the model with independent agents, with the probability of independence $p$, all four agents act independently, i.e., they do not follow either majority or minority opinion. Then, with probability $1/2$, the agents change their opinion oppositely, $\pm S(t) = \mp S(t+1)$. 
    \item For the model with anticonformist agents, with the probability of anticonformity $p$, agents change their opinion oppositely $\pm S(t) = \mp S(t+1)$ if all four agents have the same opinion (state).
    \item Otherwise, with probability $(1-p)$, each local group member follows the group's majority opinion.
\end{enumerate}

Note that, for simplicity, we use the same notation $p$ for probabilities of independence and anticonformity. To avoid ambiguity, we ensure that in all figures and discussions of each case later on we explicitly mention the context of $p$ whether it is for independence or anticonformity.

On the complete graph, we perform analytical calculations for a finite system to understand the phase transition of the model. The complete graph or fully-connected network is a topology that describes each member as neighboring and can interact with each other with the same probability \cite{nyczka2013anticonformity}. This topology is similar to the panmictic population in biology, where each member of the population can contact any other \cite{coltman2007panmictic}. In statistical physics, this topology is equivalent to the concept of the mean-field, where the state of the system only depends on the density of spin up or the order parameter of the system \cite{nyczka2013anticonformity, muslim2020phase, muslim2021phase, muslim2022opinion, sznajd2005mean, biswas2011mean}.

In the Monte Carlo (MC) simulation, the order parameter $m$ can be computed using $m = \langle \sum_i S_i/N \rangle$, where $\langle \cdots \rangle$ is the sample average. 

The system is initially disordered with zero magnetization since the density probabilities of spin-up and spin-down are set equal. Based on the microscopic interaction above, for the model with the independent agents, we generate the random number $r$, namely if $r$ is less than the probability of independence $p$, then with probability $1/2$, all four agents change their opinion oppositely. Otherwise, the agent follows the local majority opinion. The same is done for the model with the anticonformist agent, i.e., we generate the random number $r$, and if $r$ is less than the probability of anticonformity $p$, all four agents change their opinion oppositely.

To analyze the critical point of the system, we consider the Binder cumulant $U$ and the susceptibility $\chi$ that are defined as \cite{binder1981finite}:
\begin{align}
	U =& 1 - \dfrac{\langle m^4 \rangle}{3\, \langle m^2 \rangle^2}, \label{eq:binder} \\
	\chi =& N \left( \langle m^2 \rangle - \langle m \rangle^2\right), \label{eq:chi}
\end{align}
where $m$ here is the order parameter of the system.  To define the universality class of the model, we estimate the critical exponents using finite-size scaling relations~\cite{cardy1996scaling}:
\begin{align}
	m ( N) & \sim N^{-\beta/\nu}, \label{eq4} \\
	\chi(N) & \sim N^{\gamma/\nu}, \label{eq5}\\
	U(N) &\sim \text{constant}, \label{eq6}\\
	p_{c}(N)-p_{c} & \sim N^{-1/\nu}, \label{eq7}
\end{align}
which are valid near the critical points.

\section{Result and discussion}
\label{sec.3}

\subsection{Analytical results}

An analytical calculation could be performed to the complete graph model, the simplest model of networks, to obtain its order parameter.  In this section, we perform the analytical calculation for a finite system on the complete graph for four-agent interaction based on the majority rule model. The order parameter (magnetization) of the system is expressed as follows:
\begin{equation}\label{eq:mag}
     m = \dfrac{\sum_{i=1}^{N}S_i}{N} = \dfrac{N_+-N_-}{N_++N_-},
\end{equation}
where we define $N_+$  and $N_-$ as the number of spin states with up and down orientations, respectively. If we assume the probability of choosing spin-up states in the population is $k=N_+/N$, then from Eq.~\eqref{eq:mag} we have a relation $m =2\,k-1$. During the dynamics process, the spin-up (-down) increases (decreases) by $+1$ ($-1$) with probability density $\text{Pr}_+$ ($\text{Pr}_-$) or remain constant with probability of $\left(1-\text{Pr}_+-\text{Pr}_-\right)$, which can be written as:
\begin{equation}\label{eq:pr-probs}
	\begin{aligned}
		\text{Pr}_{+}  & = \text{prob}\left(k \rightarrow k+ 1/N\right), \\ 
		\text{Pr}_{-} & = \text{prob}\left(k \rightarrow k- 1/N\right). \\
	\end{aligned}
\end{equation}
%
The rate of the spin-up density $k$ is described by a master equation as follows \cite{krapivsky2010kinetic}:
\begin{equation}\label{eq:gain-loss}
    \frac{dk}{dt} = \text{Pr}_+-\text{Pr}_-.
\end{equation}
The master equation [Eq.~\eqref{eq:gain-loss}] represents the ``gain-loss" of the spins in the system during the dynamics process. Therefore, we can obtain the formula of probability $k$ by considering the equilibrium state of the system. 

\subsubsection{The model with independent agents}
For the model with independent agents, based on the majority rule model, eight configurations of agents follow the standard majority rule, i.e., agents follow the majority opinion (state) of the group, or in the socio-psychological context similar to the conformity behavior and otherwise follow the independence behavior as mentioned in the microscopic interaction above. In addition, from the configurations of agents, four configurations make spin flips from up to down state with probability $\left(1-p\right)$, and four other configurations make the spin flips from down to up with probability $\left(1-p\right)$. Therefore, the probability density of spin-up increases $\text{Pr}_{+}$ and decreases $\text{Pr}_{-}$ in Eq.~\eqref{eq:pr-probs} for all configurations can be written as:
\begin{equation}
    \begin{aligned}
    \text{Pr}_{+} & = N_{-} \left(1-p\right)\,\dfrac{4 \prod_{i=0}^{2} \left(N_{+}-i\right) }{\prod_{i=0}^{3} \left(N-i\right)}+ \dfrac{2\,N_{-}\,p}{N},\\
    \text{Pr}_{-} & = N_{+} \left(1-p\right) \dfrac{4 \prod_{i=0}^{2} \left(N_{-}-i\right) }{\prod_{i=0}^{3} \left(N-i\right)}+ \dfrac{2\,N_{+}\,p}{N}.\\
    \end{aligned}
    \label{eq:prob}
\end{equation}

\begin{figure*}[bt]
    \centering
    \includegraphics[width = 0.85\linewidth]{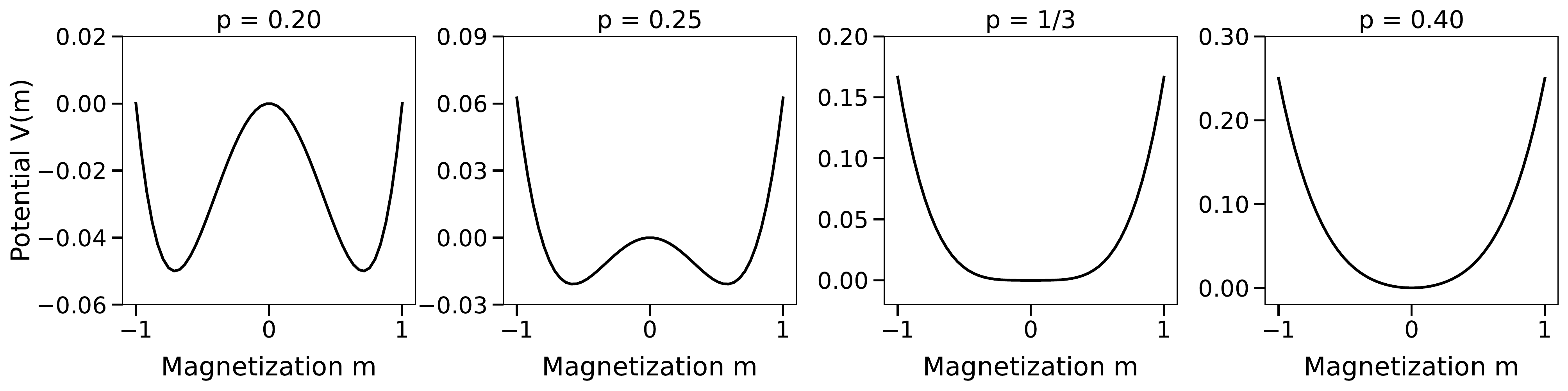}
    \caption{The bistable potential $V(m)$ (Eq.~\eqref{eq:potential}) for a large population of the majority rule model for four-agent interaction with independence agent on the complete graph. As seen, below  the probability independence $p = 1/3$ the system is in a bistable state at $m \neq 0$ and at $p = 1/3$ there is a transition from a bistable to monostable state indicates that the system undergoes a continuous phase transition at $p=p_c = 1/3$.}
    \label{fig2}
\end{figure*}

The order parameter $m$ can be obtained from the effective potential $V(m)$ of the system. The effective potential $V(m)$ can be obtained from the 'gain-loss' equation in Eq.~\eqref{eq:gain-loss}, where it can be defined as the 'effective force' $F = \text{Pr}_{+}-\text{Pr}_{-}$, is a force that make the spins flip \cite{nyczka2012opinion}.
Therefore, by integrating $F$ respects to $m$, the effective potential of the system is
\begin{equation}\label{eq:potential}
    V(m) = \dfrac{\left(1-p\right)N^2 m^4 + \left( \left(6\,p-2\right) N^2 - 12 p\,N + 8\,p\right)m^2}{4 \left(N-1\right) \left(N-2\right)}.
\end{equation}
The potential $V(m)$ is also called a bistable potential. Plot of Eq.~\eqref{eq:potential} for $N \to \infty$ and for several values of the probability independence $p$ is exhibited in Fig.~\ref{fig2}. 
Based on Fig.~\ref{fig2}, one can see that for
$p < 1/3 $ there is a bistable state at the magnetization $ -1 < m < 1$ and the
potential leads to monostable state at $m = 0$ as $p$ increases. At $p = 1/3$, there is a transition bistable-monostable state that indicates there is a phase transition. From the sociophysics point of view can be treated
as the movement of public opinion $m$ in bistable potential $V(m)$ \cite{nyczka2012opinion}. The system is very stable for small probability independence $p$, and the stability decreases as $p$ increases. This condition means that for low independence $p$, the possibility of transition between two stable opinions is low, and the stability decreases as $p$ increases. At the critical of independence $p = p_c$, no majority opinion indicating the stability at $m = 0$.

The order parameter $m$ of the system can be obtained from the extremum condition of Eq.~\eqref{eq:potential}, and for non-zero $m$, we find
\begin{equation}\label{eq:order1}
    m = \pm \sqrt{\dfrac{N^2\left(1-3p\right) + \left(6N-4\right)p}{\left(1-p\right)N^2}}.
\end{equation}
For $N \to \infty$, the order parameter $m \approx \pm \sqrt{\left(1-3\,p\right)/\left(1-p\right)}$, which is can be written as $p_c \sim (p-p_c)^{\beta}$, where $\beta = 1/2$ is a critical exponent correspond to the order parameter $m$. The critical point $p_c$ of the system can be obtained by setting $m = 0$ of Eq.~\eqref{eq:order1} namely
\begin{equation}\label{eq:criticp} 
    p_c = \dfrac{N^2}{3N^2-6N+4},
\end{equation}
which indicates that the model undergoes a continuous phase transition with the critical point at $p_c$ as in Eq.~\eqref{eq:criticp}.  As formulated in Eq.~\eqref{eq:criticp}, the critical point $p_c$ decreases as $N$
increases and reaches an asymptotic point $p_c \to 1/ 3$
for $N \to \infty $. It needs to be stressed that the model is valid for a large population ($N >> 1$) since the model is defined on the complete graph.
\subsubsection{The model with anticonformist agents}
\begin{figure*}[bt]
    \centering
    \includegraphics[width = 0.85\linewidth]{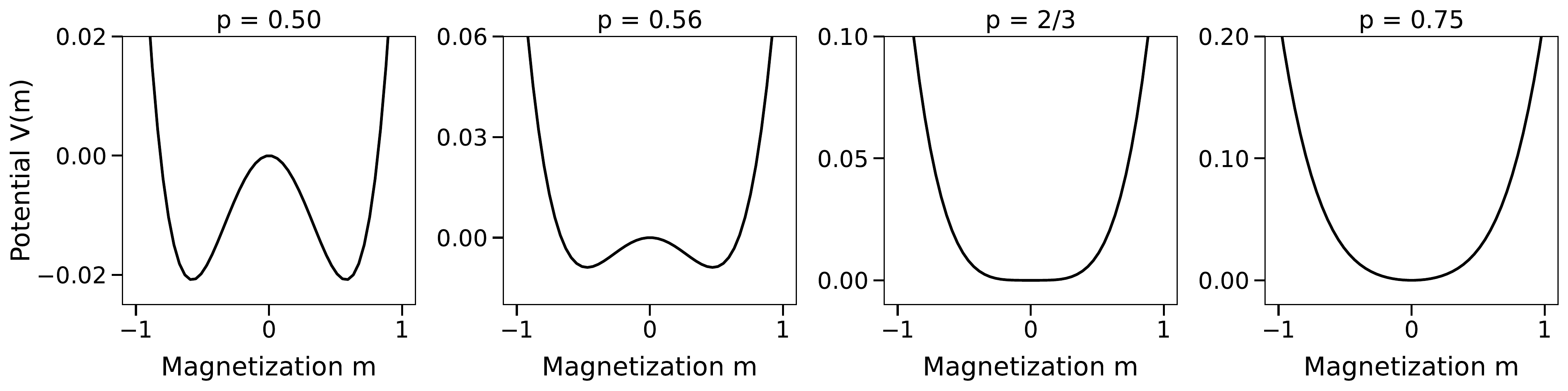}
    \caption{The bistable potential $V(m)$ (Eq.~\eqref{eq:pot_anti}) for a large population of the majority rule model for four-agent interaction with anticonformity agent on the complete graph. As seen, below  the probability anticonformity $p = 2/3$ the system is in a bistable state at $m \neq 0$ and at $p = 2/3$ there is a transition from a bistable to monostable state indicates that the system undergoes a continuous phase transition at $p=p_c = 2/3$.}
    \label{fig:pot_anti}
\end{figure*}

For the model with the anticonformist agent, from all possibles spin or agent configurations, two configurations, i.e., all spins up $++++$ and all spins down $----$ flip with probability $p$ otherwise with probability $1-p$ follow the standard majority rule model. Therefore, the probability density of spin-up increases $\text{Pr}_{+}$ and decreases $\text{Pr}_{-}$ of the model can be written as:
\begin{equation}
    \begin{aligned}
    \text{Pr}_{+} & = N_{-} \left(1-p\right)\dfrac{4 \prod_{i=0}^{2} \left(N_{+}-i\right) }{\prod_{i=0}^{3} \left(N-i\right)}+ \dfrac{\prod_{i=0}^{3}\left(N_{-}-i\right)\,p}{\prod_{i=0}^{3} \left(N-i\right)},\\
    \text{Pr}_{-} & = N_{+} \left(1-p\right) \dfrac{4 \prod_{i=0}^{2} \left(N_{-}-i\right) }{\prod_{i=0}^{3} \left(N-i\right)}+ \dfrac{\prod_{i=0}^{3}\left(N_{+}-i\right)\,p}{\prod_{i=0}^{3} \left(N-i\right)}.\\
    \end{aligned}
    \label{eq:prob_indep}
\end{equation}
In the same way as the model with the independent agent case above, by using the effective force $F$ and integrating it for the parameter order $m$, the effective potential $V(m)$ of the model is
\begin{equation}\label{eq:pot_anti}
    V(m) = \dfrac{\left(2-p\right)\, N^2\,m^4 + \left(\left(6\,p-4\right)\, N^2 - 12\,p\,N+8 \right)\, m^2}{8\left(N-2\right)\left(N-1\right)}.
\end{equation}
One can see there is a bistable state below the critical point $p_c = 2/3$ and the system toward to monostable state as $p$ increases, as  shown in Fig.~\ref{fig:pot_anti}.
For the extremum condition of Eq.~\eqref{eq:pot_anti}, the order parameter for non zero $m$ of the model is
\begin{equation}\label{eq:order_antconfor}
    m = \pm \sqrt{\dfrac{\left(2-3\,p\right)\, N^2+\left(6\,N-4\right)\,p}{\left(2-p\right)\,N^2}},
\end{equation}
and we find the critical point $p_c$ of the model, namely
\begin{equation}\label{eq:critical_anti}
    p_c = \dfrac{2 N^{2}}{3 N^{2}-6 N+4}.
\end{equation}
For a large population $N >> 1$, the critical point is $p_c \approx 2/3$. The order parameter $m$ of this model is also can be written as $m \sim (p-p_c)^{\beta}$, where $\beta  = 1/2$. One can see that the model undergoes a continuous phase transition with the critical anticonformity $p_c$ given by Eq.~\eqref{eq:critical_anti}.

One can compare the result of the model with independent agents versus the model with anticonformist agents, that both models undergo a continuous phase transition with a different critical point. The critical point of the model with the independent agent is smaller than the model with the anticonformist agent, as exhibited in Fig.~\ref{fig:order_complete}. This condition is caused by the effect of independence behavior on the system has a more significant 'noise effect' than anticonformity behavior. Therefore, the model with the independent agent has a higher probability of phase changing. From a sociophysics point of view, the model with the independent agent has a higher chance reach a stalemate situation than the model with the anticonformist agent.

\subsubsection{Probability density function}
In this subsection, we analyze the model on compete graph from the probability density function of the order parameter $m$. We start from the master equation that describes the probability density function of the spin-up density $k$ at time $t$ such as:

\begin{align}\label{eq:density}
    \text{P}\left(k,t+\Delta_t \right) = \,& \text{Pr}_+ \left(k-\Delta_N \right) \text{P}\left(k-\Delta_N, t \right) \nonumber \\& + \text{Pr}_- \left(k+\Delta_N \right) \text{P}\left(k+\Delta_N, t \right) \nonumber \\
    & + \left(1-\text{Pr}_+-\text{Pr}_-\right) \text{P}\left(k,t\right).
\end{align}
For a large population $N$, we can approximate the master equation in Eq.~\eqref{eq:density} using the Fokker-Planck equation as follows \cite{frank2005nonlinear}:

\begin{equation}\label{eq:fokker-planck}
	\dfrac{\partial}{\partial t} \text{P}(m,t) = \dfrac{4}{N}\dfrac{\partial^2}{\partial m^2} \left[\gamma_1 \text{P}(m,t)\right] -2 \dfrac{\partial }{\partial m} \left[\gamma_2 \text{P}(m,t)\right],
\end{equation}
which describes the time evolution of the probability density function $\text{P}(m)$ of the order parameter $m$. The parameters $\gamma_1$ and $\gamma_2$ are the diffusion-like and drift-like coefficients depending on the magnetization $m$, respectively, defined as:
\begin{equation}
  \begin{aligned}
    2\,\gamma_1 = & \left[ \text{Pr}_+ + \text{Pr}_- \right], \\
    \gamma_2 = & \left[ \text{Pr}_+ - \text{Pr}_- \right],
\end{aligned}  
\end{equation}
where $\text{Pr}_+$ and $\text{Pr}_-$ depend on the model. To analyze the phase transition of the model, we consider the stationary condition of Eq.~\eqref{eq:fokker-planck}, which gives the general solution as follows:

\begin{equation}\label{fokker-planck_solution}
	\text{P}(m)_{st} \sim \gamma_1^{-1} \exp\int \dfrac{N\gamma_2}{2\, \gamma_1} dm.
\end{equation}

The diffusion-like $\gamma_1$ and drift-like coefficients $\gamma_2$ of the model with independent agents for a large population are given by:
\begin{equation}
 \begin{aligned}
    \gamma_1 \approx & \left[\left(p-1 \right) m^{4}+3\,p+1 \right]/4, \\
    \gamma_2 \approx & \left[\left(p-1 \right) m^{2}-3\,p +1\right] m.
\end{aligned}   
\end{equation}
Therefore, the general solution of Eq.~\eqref{fokker-planck_solution} is given by:
\begin{align} \label{eq:Prob_indep}
    	\text{P}(m)_{st} \sim \,& \gamma_1^{-1} \nonumber \\
    	& \times \exp \left( \frac{N}{2}\, \left( \ln \left(4\gamma_1\right)+ \dfrac{\left(2-6\,p \right)\arctan \sqrt{\dfrac{p-1}{3p+1}}m^2}{\sqrt{3\,p^2-2\,p-1}}\right) \right).
\end{align}

\begin{figure*}[bt]
    \centering
    \includegraphics[width = 0.8\linewidth]{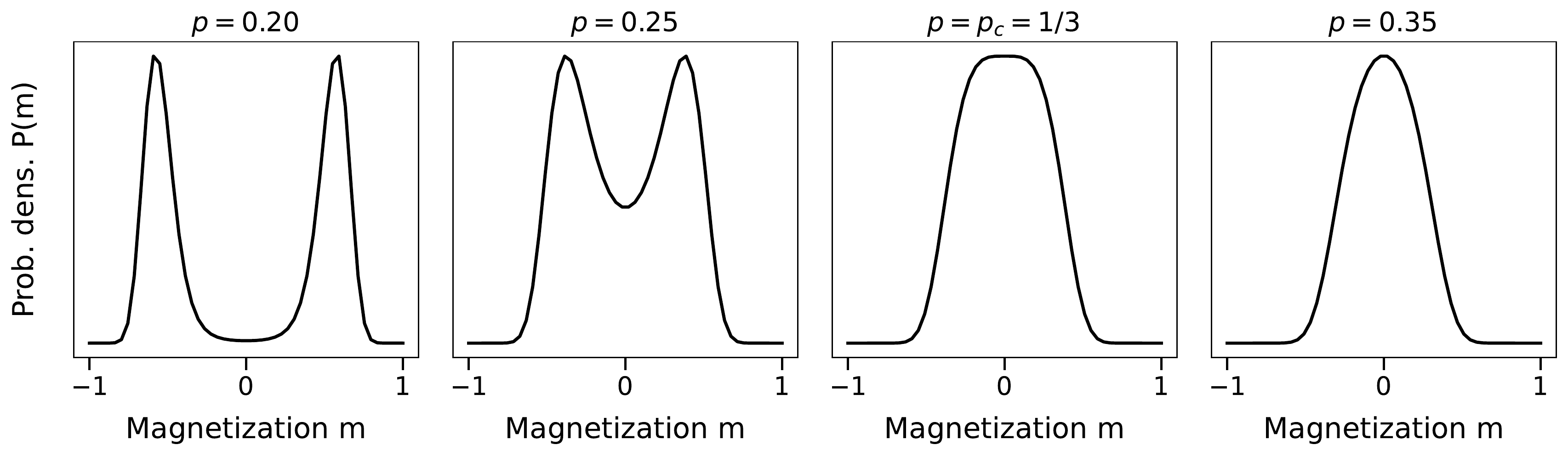}
    \caption{Stationary probability density function $\text{P}(m)_{st}$ in Eq.~\eqref{eq:Prob_indep} (not normalized) of the order parameter $m$ of the model with independence agent for typical values of independence $p$ and the population size $N = 200$. One can see there are two 'peaks' for $p < p_c = 1/3$ at $m(p,N) \neq 0$ indicates there is majority opinion in the system, and reach one peak for $p >p_c$ at $m = 0$ indicates there is no majority opinion in the system. The peaks at $m = \pm 1$ for the probability independence $p = 0$ (the system without independence agents). The peaks are getting lower as $p$ increases and approach each other. This phenomenon indicates that the model undergoes a continuous phase transition.}
    \label{fig:prob_indep}
\end{figure*}

\begin{figure*}[t!]
    \centering
    \includegraphics[width = 0.8\linewidth]{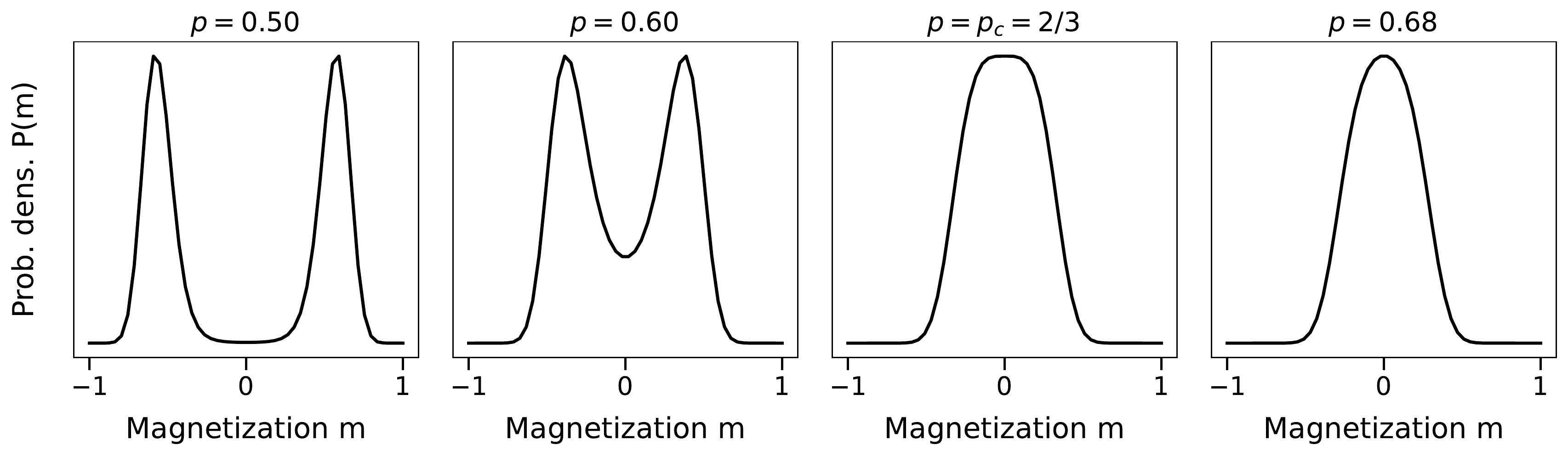}
    \caption{Stationary probability density function $\text{P}(m)_{st}$ in Eq.~\eqref{eq:Prob_anti} (not normalized) of the order parameter $m$ of the model with anticonformity agents for typical values of anticonformity $p$ and the population size $N = 200$. One can see there are two 'peaks' for $p < p_c = 2/3$ at $m(p,N) \neq 0$ indicates there is majority opinion in the system, and reach one peak for $p > p_c$ at $m = 0$ indicates there is no majority opinion in the system. The peaks at $m = \pm 1$ for the probability anticonformity $p = 0$ (the system without anticonformist agents). The peaks are getting lower as $p$ increases and approach each other. This phenomenon indicates that the model undergoes a continuous phase transition.}
    \label{fig:prob_anti}
\end{figure*}

The diffusion-like $\gamma_1$ and drift-like coefficients $\gamma_2$ of the model with an anticonformist agent for a large population are given by:
\begin{equation}
 \begin{aligned}
    \gamma_1 \approx & \left[\left(5 p -4\right) m^{4}+6\, p \,m^{2}-3\,p+4 \right]/16, \\
    \gamma_2 \approx & \left[\left(p-2 \right) m^{3}- \left(3\,p +2 \right)m\right]/2,
\end{aligned}   
\end{equation} with the general solution of Eq.~\eqref{fokker-planck_solution} for this model is given by:

\begin{align} \label{eq:Prob_anti}
    	\text{P}(m)_{st} \sim \, & \gamma_1^{-1} \exp \left({\dfrac{N \ln \left(16\,\gamma_1 \right) \left(p-2\right)}{\left(5 p -4\right)}} \right) \nonumber \\
    	& \times \exp \left({\dfrac{N \xi_2\, \mathrm{arctanh}\left(\dfrac{\left(5\,p-4\right) m^{2}+3\,p}{2\, \xi_1}\right)}{\xi_1\, \left(5 p -4\right)}} \right),
\end{align}
where $\xi_1 = \sqrt{6 p^{2}-8 p +4}$ and $\xi_2 = \left(18\,p^2 - 28\, p + 8 \right)$. Plot of Eqs.~\eqref{eq:Prob_indep} and \eqref{eq:Prob_anti}  for typical values of probability independence $p$ for the model with independent agents and for typical values of anticonformity $p$ for the model with anticonformist agents are exhibited in Figs.~\ref{fig:prob_indep} and \ref{fig:prob_anti}, respectively. One can see that for both models, there are two peaks or maxima of $\text{P}(m)_{st}$ at the magnetization $m(p,N) \neq 0 $ for $ p $ below the critical point $p_c$  and reach one peak at the magnetization $m(p,N) = 0$ as $p$ increases. For the case without independence or anticonformity agents ($p = 0$), both systems are in a homogeneous state with all spin-up or all spin-down (ferromagnetic or complete consensus), indicated by two peaks of $\text{P}(m)_{st}$ at $m =\pm 1$. The peaks of  $\text{P}(m)_{st}$ are getting lower and more narrow as the $p$ increases. One can also obtain the order parameter $m$ from the probability density function by considering the extremum condition of Eq.~\eqref{eq:prob_indep} and will be given the same result as in Eq.~\eqref{eq:order1} for the model with independent agents and Eq.~\eqref{eq:order_antconfor} for the model with anticonformist agent.

\subsection{Monte Carlo Simulation}

\subsubsection{The model on the complete graph}
\label{subsec:the model on complete graph}
\begin{figure}[bt]
    \centering
     \includegraphics[width = \linewidth]{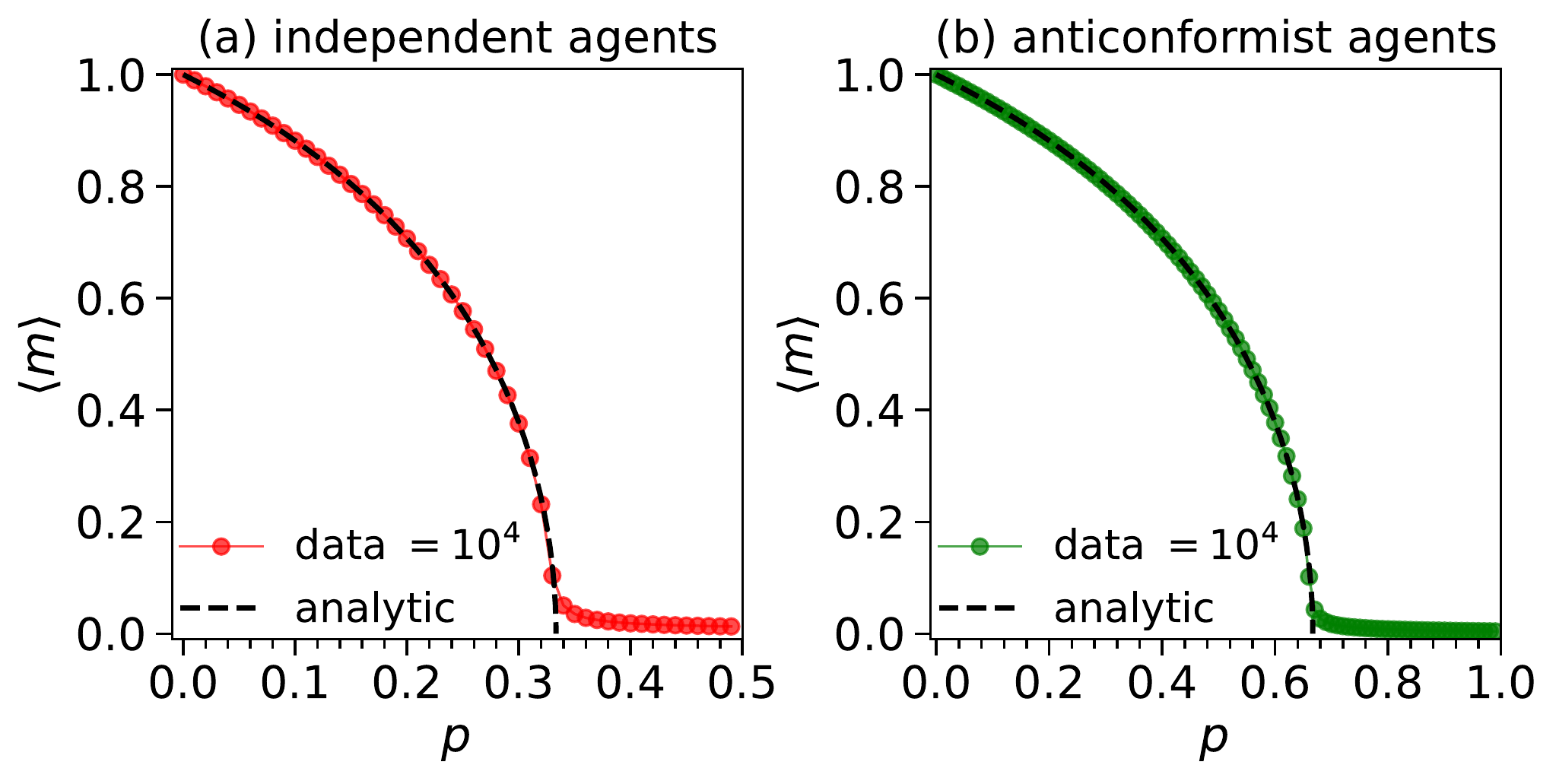}
    \caption{Numerical results for the order parameter $m$ versus probability $p$ of independence or anticonformity for the model with independent and anticonformist agents defined on the complete graph. It can be seen that both models undergo a continuous phase transition and the agreement between numerical and analytical result in Eq.~\eqref{eq:order1} for independence agent [panel (a)], and Eq.~\eqref{eq:order_antconfor} for anticonformist agent [panel (b)].The population size $N = 10000$. }
    \label{fig:order_complete}
\end{figure}

In order to verify the analytical result in the previous section, we perform Monte Carlo (MC) simulations to obtain the critical point and define the universality class of the model. As seen in Eqs.~\eqref{eq:order1} and \eqref{eq:order_antconfor}, the parameter control of the system is only the probability of the independence or anticonformity $p$. 
We here consider a large population size $N = 10000$, and each point averages over $10000$ samples. The comparison between the analytical result in Eqs.~\eqref{eq:order1} and \eqref{eq:order_antconfor}, and the numerical result for the order parameter $m$ is exhibited in Fig.~\ref{fig:order_complete}. One can see the agreement of the results and show that the model undergoes a continuous phase transition for both the models with independent agents (a) and with anticonformist agents (b).  

\begin{figure*}[t!]
    \centering
    \includegraphics[width = 0.95\linewidth]{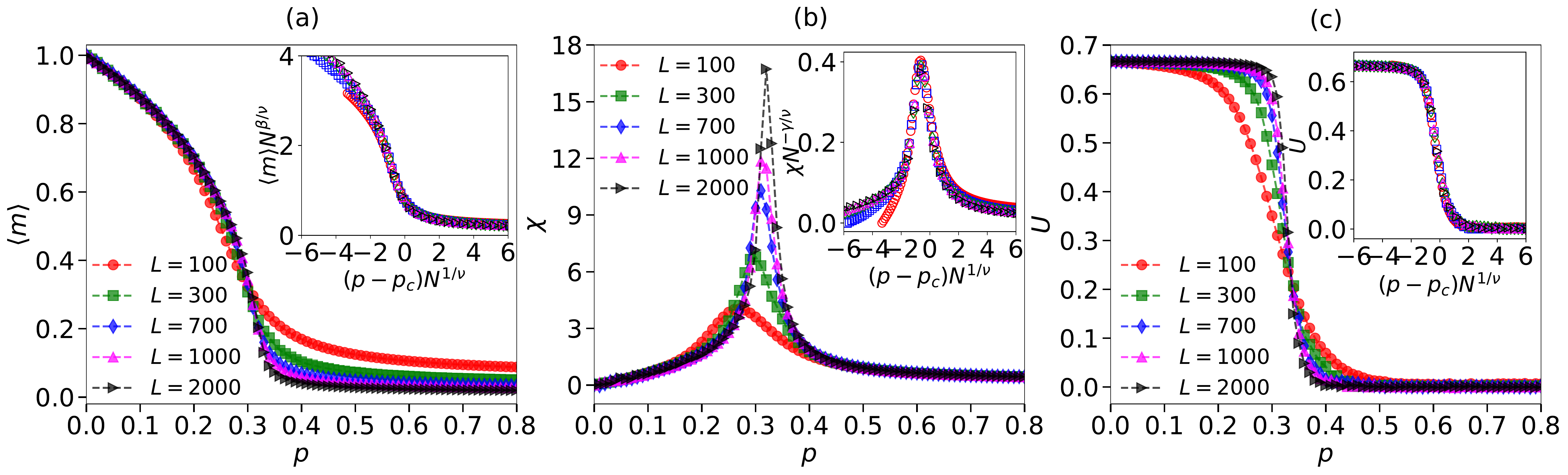}
    \caption{Results of numerical calculations of (a) order parameter $m$, (b) susceptibility $\chi$, and (c) Binder cumulant $U$ as a function of independence probability $p$. Based on the finite-size scaling analysis, we can obtain the critical point $p_c$ from the crossing of lines of Binder cumulant $U$ that occurrs at $p_c \approx 0.334$ [see panel (c)] and the critical exponents $\nu \approx 2, \beta \approx 0.5,$ and $\gamma \approx 1.0$ that makes the best collapse of all $N$ (see the insets). The critical point agrees with the analytical results in Eq.~\eqref{eq:criticp} for a large $N$ that is $p_c \approx 1/3$ since the model is defined on the complete graph. }
    \label{fig:complete_graph_indep}
\end{figure*}

\begin{figure}[bt]
    \centering
    \includegraphics[width = \linewidth]{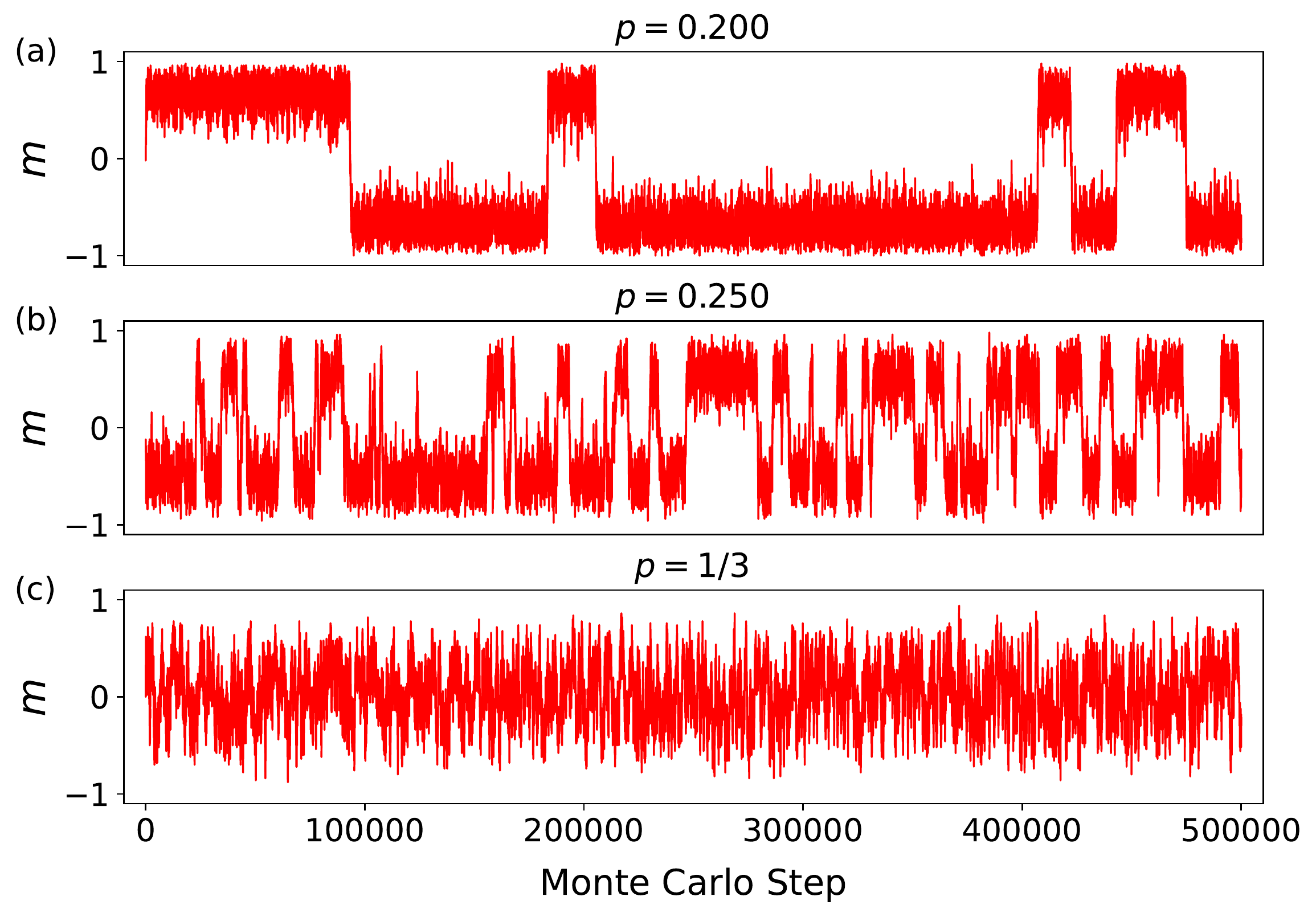}
    \caption{The evolution of the magnetization $m$ per site over the MC step of the model with independent agents. As seen, for the probability of the independence $p$ below the critical point $p_c =1/3$, there is bistable state that indicated by value of the magnetization $m \to \pm 1$, and for $p = p_c = 1/3$, the system toward to monostable state at $m \to 0$. This phenomenon indicates a continuous phase transition. The population $N = 100$.}
    \label{fig:mcs_indep}
\end{figure}

To obtain critical points of the system, we consider several population sizes $N = 100, 300, 700, 1000, 2000$. Based on the finite-size scaling relations in Eqs.~\eqref{eq4}--\eqref{eq7}, the critical point of the model is obtained from the crossing of lines of the Binder cumulant $U$ that occurs at $p_c \approx 0.334$, as depicted in Fig.~\ref{fig:complete_graph_indep}(c). The value of this critical point agrees with the analytical result in Eq.~\eqref{eq:criticp} for $N \to \infty$ since the model is valid for a large population. We also obtain the critical exponents $\beta, \gamma, $ and $\nu$ namely $\beta \approx 0.5, \gamma\approx 1.0, $ and $\nu \approx 2.0 $ that make the best collapse of all data as exhibited in the inset of Fig.~\ref{fig:complete_graph_indep}. These critical exponents indicate the model is in the same universality class as that of the Sznajd model with contrarian and independence agent \cite{muslim2022opinion}, the kinetic exchange opinion model with two-one agent interactions \cite{biswas2011mean, biswas2012disorder, crokidakis2014phase}, and the mean-field Ising model.
We also analyze the evolution of the order parameter $m$ in the MC step as exhibited in Fig.~\ref{fig:mcs_indep}. One can see that for low independence $p < p_c$, there is a fluctuation of two bistable states indicating the model undergoes a continuous phase transition. The system goes to monostable for $p$ increases, which agrees with the description of bistable potential in the analytical result (see Fig.~\eqref{fig2}).

\begin{figure*}[t!]
    \centering
    \includegraphics[width = 0.95\linewidth]{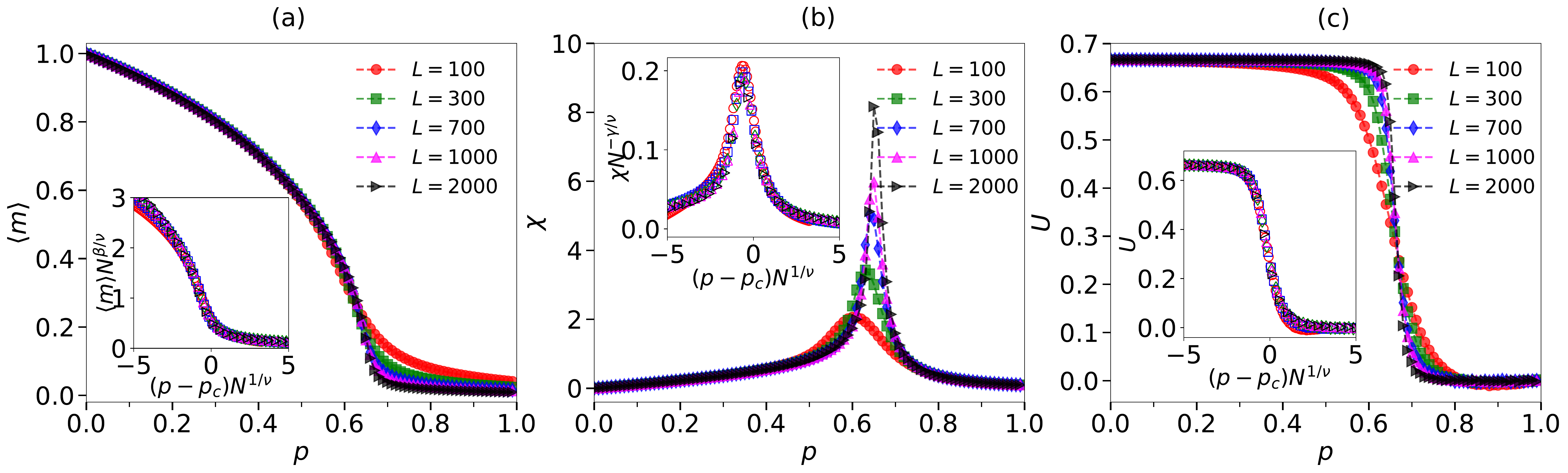}
    \caption{Results of numerical calculations of (a) order parameter $m$, (b) susceptibility $\chi$, and (c) Binder cumulant $U$ as a function of anticonformity probability $p$. Based on the finite-size scaling analysis, we can obtain the critical point $p_c$ from the crossing of lines of Binder cumulant $U$ that occurs at $p_c \approx 0.667$ [see panel (c)] and the critical exponents $\nu \approx 2, \beta \approx 0.5,$ and $\gamma \approx 1.0$ that make the best collapse of all $N$ (see the insets). The critical point agrees with the analytical results in Eq.~\eqref{eq:critical_anti} for a large $N$ that is $p_c \approx 2/3$ since the model is defined on the complete graph.}
    \label{fig:complete_graph_anti}
\end{figure*}

\begin{figure}[t!]
    \centering
    \includegraphics[width = \linewidth]{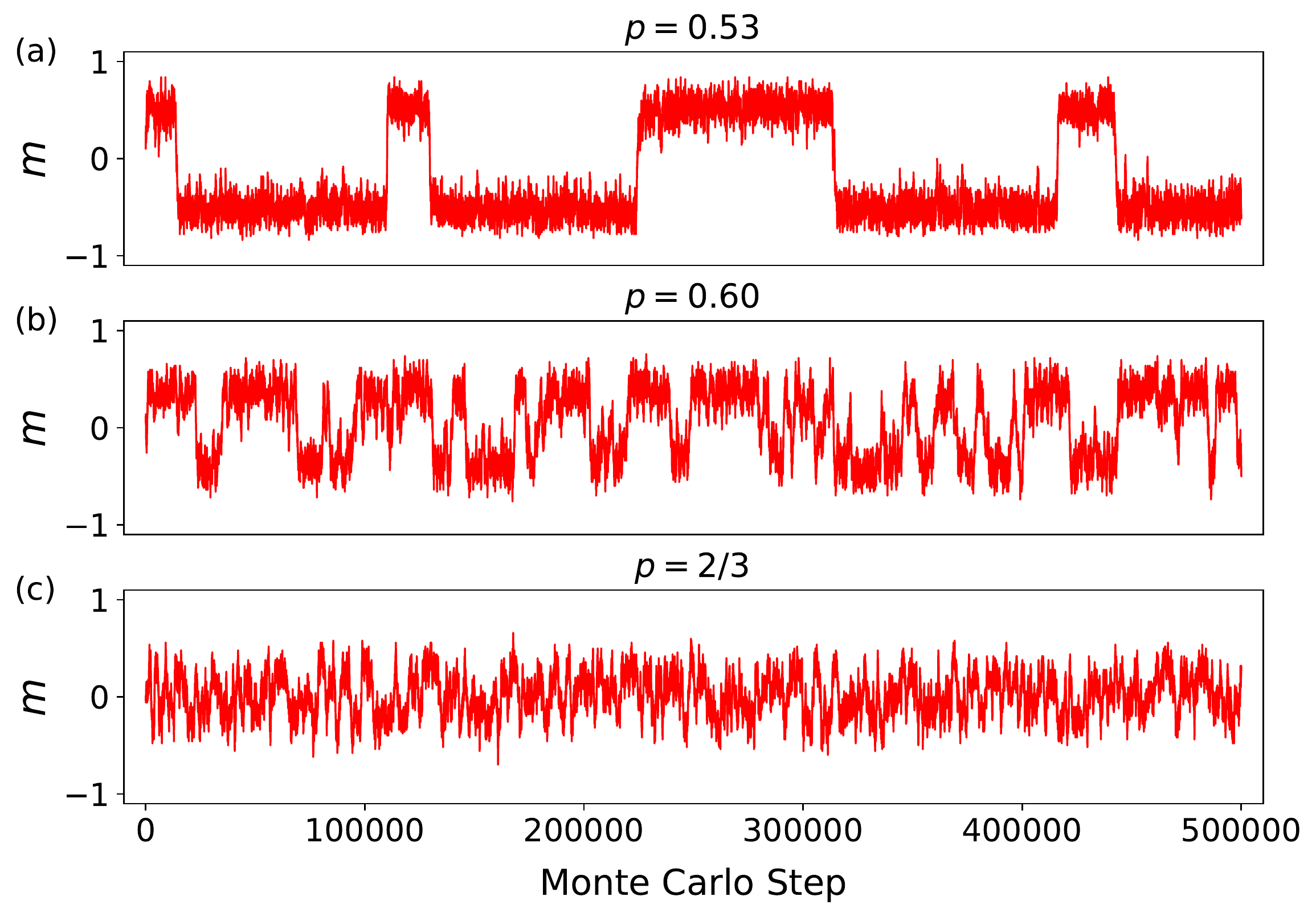}
    \caption{The evolution of the magnetization $m$ per site over the MC step for four-agent interaction. It can be seen that for the anticonformity probability $p$ below the critical point $p_c =2/3$, there is a bistable state, and for $p = p_c = 2/3$, the system toward monostable at $m \to 0$. This phenomenon indicates a continuous phase transition. The population $N = 100$.}
    \label{fig:mcs_anti}
\end{figure}

The critical point of the model with anticonformist agent occurred at $p = p_c \approx 0.667$ as exhibited in Fig.~\ref{fig:complete_graph_anti} (c), which indicates the model undergoes a continuous phase transition. This result confirms the analytical result in Eq.~\eqref{eq:critical_anti} for a large population $N$, $p_c \to 2/3$. Based on the finite-size scaling analysis, the critical exponents of the model are $\nu \approx 2.0, \beta \approx 0.5,$ and $\gamma \approx 1.0$ that make the best collapse of all data as exhibited in the inset of  Fig.~\ref{fig:complete_graph_anti}. These results indicate that the model is identical to the model with independent agents and in the same universality class as the mean-field Ising model. The evolution of the order parameter $m$ in the MC step is exhibited in Fig.~\ref{fig:mcs_anti}. One can see a similar phenomenon as in the model with the independent agent, i.e., below the critical point of the anticonformity probability $p < p_c$, there are two bistable states indicated by $m \to \pm 1$ and the system reaches a monostable state as $p$ increases.

\subsubsection{The model on the 1D chain}

\begin{figure}[t!]
    \centering
    \includegraphics[width = 8 cm]{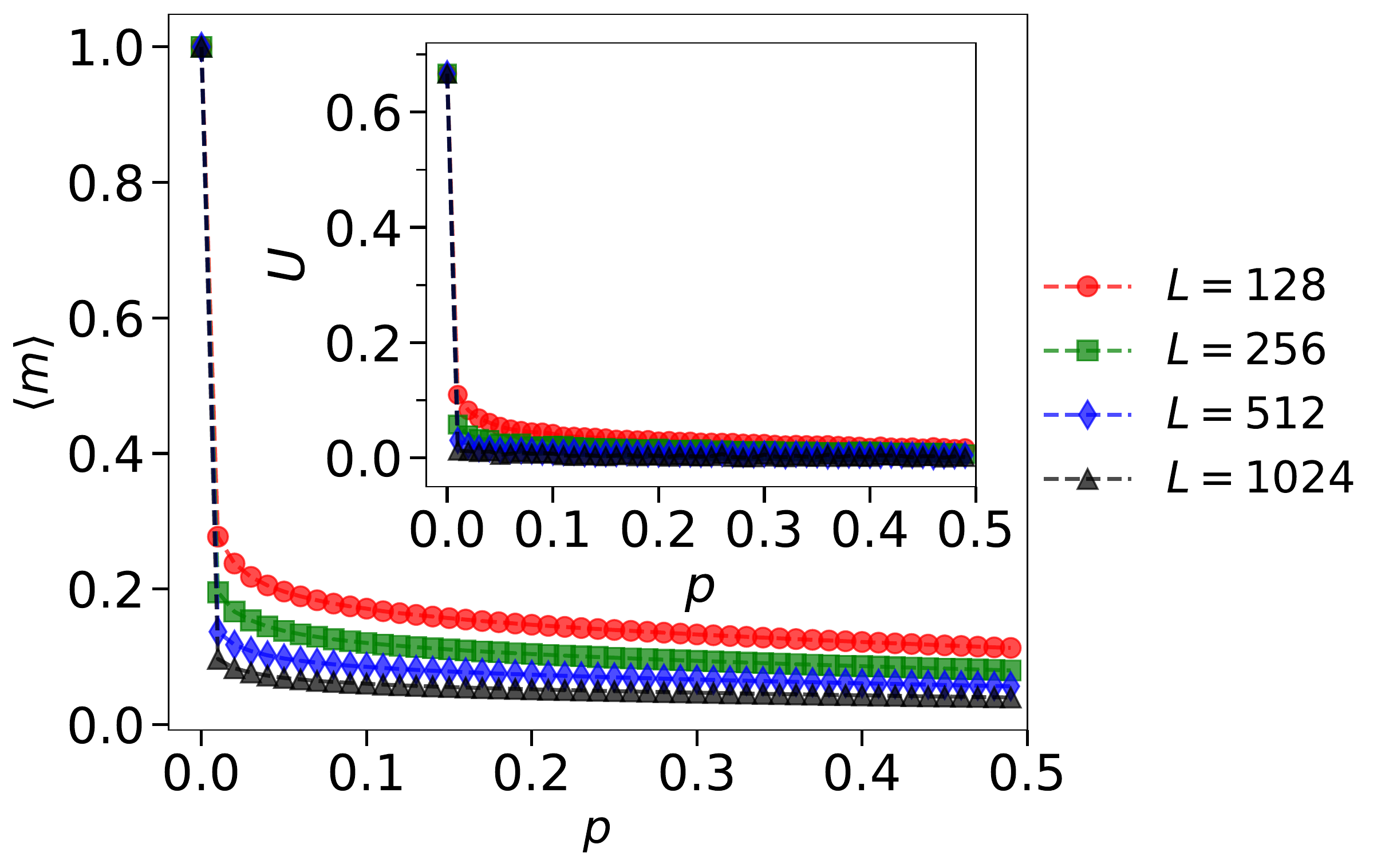}
    \caption{(Main graph) The numerical result of the order parameter $m$ as a function of the independence probability $p$ of the model with independent agents on the 1D chain. As seen in the inset graph, no crossing of lines of Binder cumulant $U$ indicates no phase transition in this model.}
    \label{fig:ring}
\end{figure}

For the model on the 1D, we consider several size of populations $N = 128, 256, 512,$ and $1024$ with periodic boundary condition. Our numerical result for the order parameter $m$ and Binder cumulant $U$ for the model with independent agents is exhibited in Fig.~\ref{fig:ring}. One can see that all of the data fall for the probability independence $p \neq 0$, and there is no crossing of lines of the Binder cumulant $U$ as shown in the inset graph. This result indicates there is no phase transition occurred in this model.

\begin{figure}[t!]
    \centering
    \includegraphics[width = 8 cm]{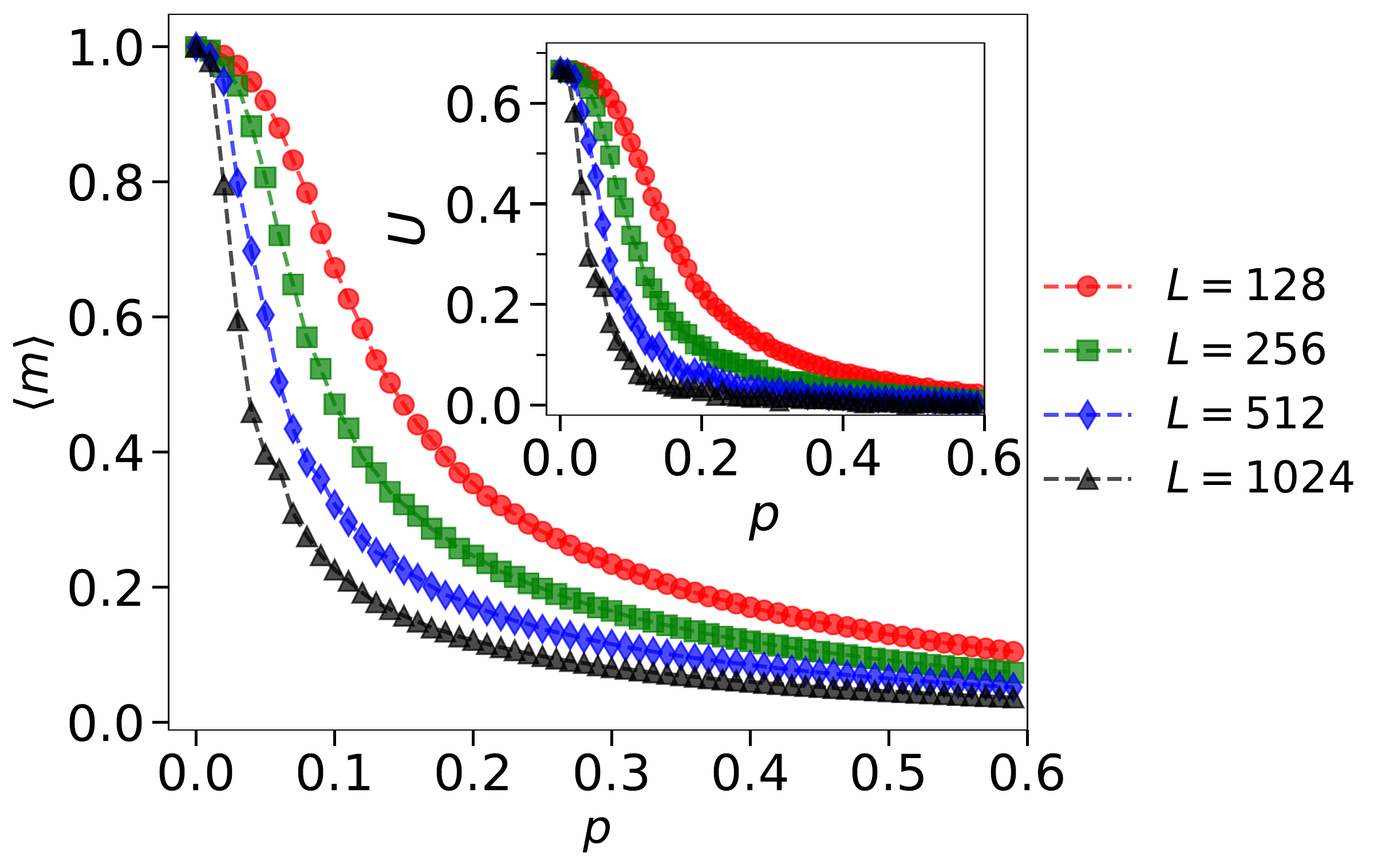}    
    \caption{(Main graph) The numerical result of the order parameter $m$ as a function of the anticonformity probability $p$ of the model with anticonformist agents on the 1D chain. As seen in the inset graph, no crossing of lines of Binder cumulant $U$ indicates no phase transition in this model.}
    \label{fig:ring_anti}
\end{figure}

The same phenomenon is observed in the model with anticonformist agents: there is no crossing of lines of the Binder cumulant $U$ as exhibited in the inset of Fig.~\ref{fig:ring_anti}, which indicates no phase transition in this model. This phenomenon also occurred in one dimensional Ising model, the 1D Sznajd model with independence \cite{sznajd2011phase}, and the 1D majority rule model with independent agents for three spins interaction \cite{crokidakis2015inflexibility}.

\subsubsection{The model on the square lattice}
\begin{figure*}[t!]
    \centering
    \includegraphics[width = 0.95\linewidth]{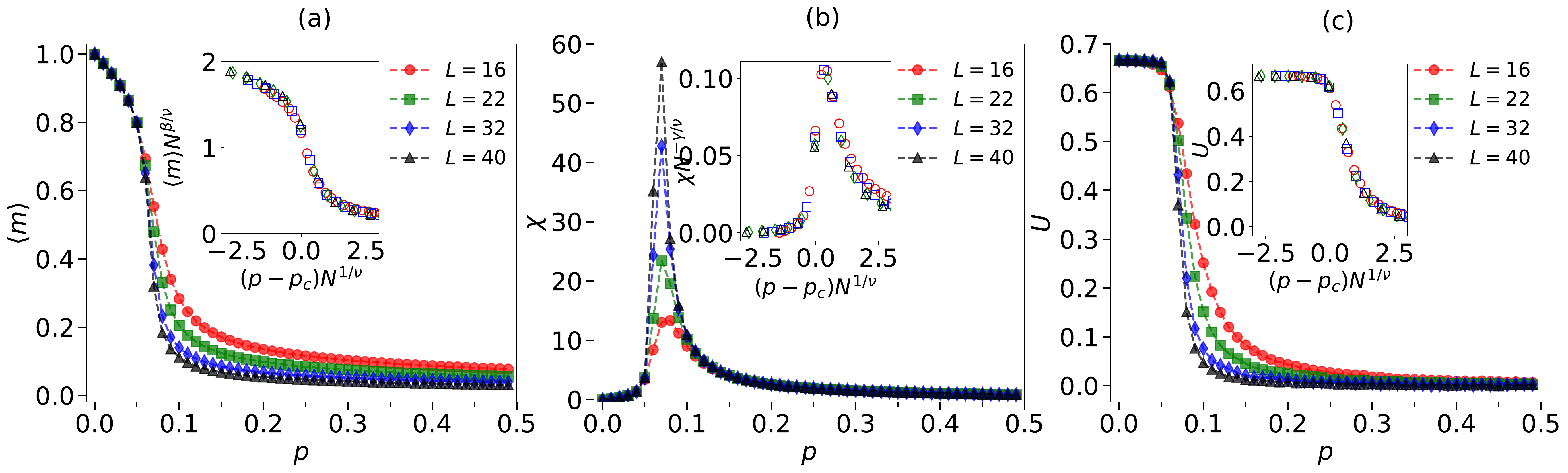}
    \caption{Results of numerical calculations of (a) order parameter $m$, (b) susceptibility $\chi$, and (c) Binder cumulant $U$ as a function of independence probability $p$ on the model on the 2D square lattice. Based on the finite-size scaling analysis, we can obtain the critical point $p_c$ from the crossing of lines of Binder cumulant $U$ that occurs at $p_c \approx 0.0608$ [see panel (c)] and the critical exponents $\nu \approx 1.75, \beta \approx 0.165, $ and $\gamma \approx 1.53$ that make the best collapse of all $N$ (see the inset graphs).}
    \label{fig:square_lattice}
\end{figure*}

For the model on the square lattice, the population size is $N = L\times L$ with $L = 16, \,22, \,32,$ and $40$ is a lattice width. The numerical results for the order parameter $m$, Binder cumulant $U$, and susceptibility $\chi$ of the model with independent agents are depicted in Fig.~\ref{fig:square_lattice}. Based on the finite-size scaling relations in Eqs.~\eqref{eq4}-\eqref{eq7}, one can see the crossing of lines of the Binder cumulant $U$ that occurred at $p_c \approx 0.0608$ indicates that the model undergoes a continuous phase transition as exhibited in Fig.~\eqref{fig:square_lattice} (c). We also estimate the critical exponents of the model using finite-size scaling, namely $\nu \approx 1.75, \beta \approx 0.16,$ and $\gamma \approx 1.53$ that make the best collapse of all the data as exhibited in the inset of graphs in Fig.~\eqref{fig:square_lattice}. The critical exponents are not identical to the 2D Ising model, which indicates they are not identical in the universality class. 

\begin{figure*}[t!]
    \centering
    \includegraphics[width = 18 cm]{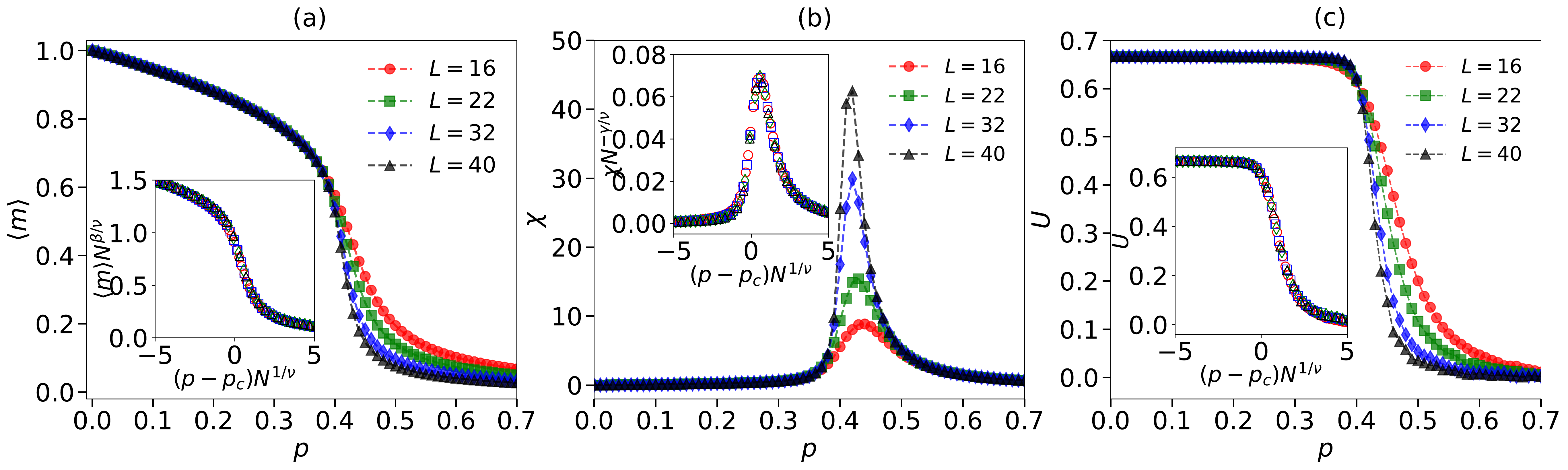}
    \caption{Results of numerical calculations of (a) order parameter $m$, (b) susceptibility $\chi$, and (c) Binder cumulant $U$ as a function of anticonformity probability $p$ on the model on the 2D square lattice. Based on the finite-size scaling analysis, we can obtain the critical point $p_c$ from the crossing of lines of Binder cumulant $U$ that occurs at $p_c \approx 0.4035$ [see panel (c)] and the critical exponents $\nu \approx 2.0 , \beta \approx 0.165 , $ and $\gamma \approx 2.75 $ that makes the best collapse of all $N$ (see the inset graphs).}
    \label{fig:square_lattice_anti}
\end{figure*}

\begin{figure}[t!]
    \centering
    \includegraphics[width = 0.95\linewidth]{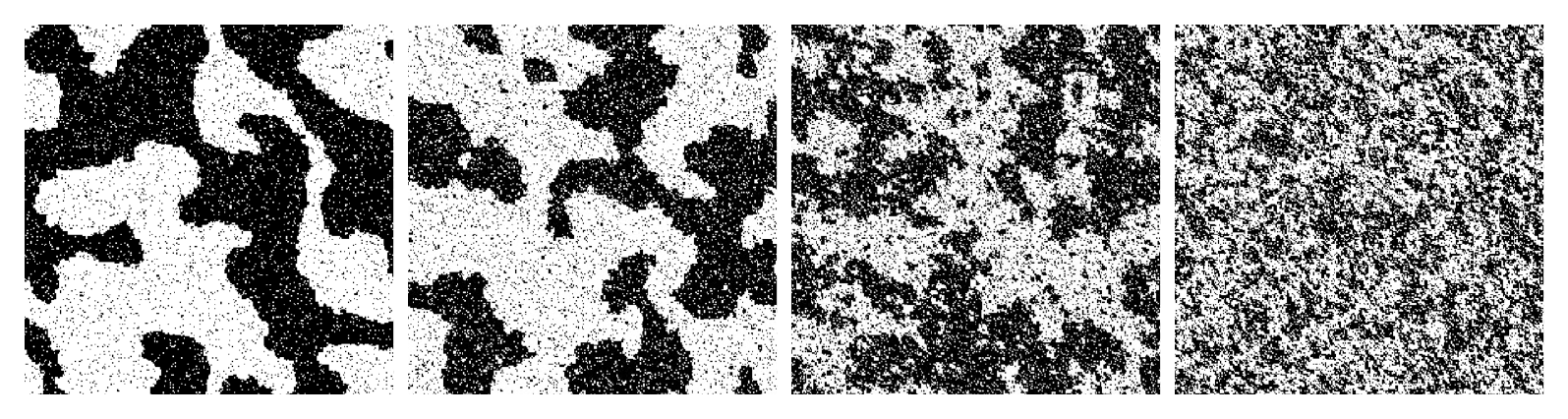}
    \caption{Snapshot of the dynamics of agents' interaction in an equilibrium state (over $500$ Monte Carlo sweep) of the model with anticonformist agents on the 2D square lattice for typical probability anticonformity $p$. From left to right $p = 0.2, 0.3, 0.4035,$ and $0.50$ respectively. The width size $L = 300$.}
    \label{fig:square_sc}
\end{figure}

The critical point of the model with anticonformist agent is $p_c\approx 0.4035$ as exhibited in Fig.~\ref{fig:square_lattice_anti} (c). We also obtain the critical exponents $\nu \approx 2.0, \beta \approx 0.165, $ and $\gamma \approx 2.75$ that make the best collapse of all data. One can see that this model also undergoes a continuous phase transition. In addition, we exhibit in Fig.~\ref{fig:square_sc} the visual of the agents interaction in equilibrium state for typical values of the probability anticonformity $ p = 0.2, 0.3, 0.4035,$ and $0.50$, respectively.

The critical exponents of the model with independent agents and the model with anticonformist agents are not the same, indicating that both models are not identical even though both models are defined on the same graph, that is, the 2D square lattice. Therefore both models are not in the same universality class, as well as the 2D majority rule model with three agents interaction \cite{crokidakis2015inflexibility}, 2D continuous opinion dynamics model \cite{mukherjee2016disorder}, 2D Sznajd model \cite{calvelli2019phase}, and 2D Ising model. This result is different from the model defined on the complete graph, i.e., both models have the same critical exponents $\nu \approx 2.0, \, \beta \approx 0.5,$ and $\gamma \approx 1.0$, which indicates the model is in the same universality class.

\section{Summary and outlook}

In this work, we investigate the majority-rule model with a nonconformist agent defined on the complete graph, 1D chain, and 2D square lattice. The nonconformist agent is divided into two types: independent and anticonformist agents. Each group member consists of four agents, they interact with each other based on the majority rule. For the model with the independent agent, with probability $p$, the agent acts independently, and further with probability $1/2$, the agent changes its opinion. Otherwise, the agent follows the standard majority rule, i.e., follow the majority opinion of the group. For the model with anticonformist agents, if all four agents have the same opinion, then with probability $p$, they change their opinion. 

Both analytical calculations and numerical simulations are performed for the model on the complete graph.
The results obtained by analytical calculations are in good agreement with those of numerical simulations. Both show that the model undergoes a continuous phase transition. 
The critical points of the model with independent and anticonformist agents are $p_c \approx 0.334$ and $p_c \approx 0.667$ respectively. Based on the finite-size scaling relations, both models have the same value of critical exponents, namely $\nu \approx 2.0$, $\beta \approx 0.5$, and $\gamma \approx 1.0$. These results indicate that both models are identical and in the same universality class as mean-field Ising model. On the 1D chain, no phase transition is observed for both models with independence and anticonformity agents. This phenomenon also occurs in the one-dimension Ising model, the 1D Sznajd model, and the majority rule model, defined on the 1D chain for the interaction of three agents. 

The continuous phase transition is observed in the model that is defined on the square lattice. The critical points of the model with independent and conformist agents are $p_c \approx 0.0608$ and $p_c \approx  0.4035$, respectively. By using finite-size scaling analysis, we also obtain the critical exponents that make the best collapse of all data, namely $\nu \approx 1.75$, $\beta \approx 0.165,$ and $\gamma \approx 1.53$ for the model with independent agents, and $\nu \approx 2.0, \beta \approx 0.165, $ and $\gamma \approx 2.75$ for the model with anticonformist agents. List of the critical point and the critical exponents of the model on all considered graphs, unless the 1D chain graph, are provided in Table~\ref{tab:table1}. The same critical exponents' values can be found for an arbitrary population size $N$, meaning that these values are universal.

\begin{table}[bt]
\caption{The critical probability independence or anticonformity $p_c$ and critical exponents $\nu, \beta$, and $\gamma$ of the majority rule model with independence agent or the model with anticonformity agent for four-agents interaction that is defined on distinct networks.}
\vspace{0.25cm}
\centering
\begin{tabular}{llcccc}
\hline
Behavior                                  & Graph             & $p_c$ & $\nu$ & $\beta$ & $\gamma$ \\ \hline
\multicolumn{1}{l}{\multirow{3}{*}{Indep.}} & C-G    & $0.334$   & $2.0$  & $0.5$  & $1.0$  \\
\multicolumn{1}{l}{}                              & 1D chain          &  $-$  & $-$   & $-$   & $-$   \\
\multicolumn{1}{l}{}                              & 2D S-L & $0.0608$   & $1.75$  & $0.165$   & $1.53$   \\ \hline
\multirow{3}{*}{Anticonf.}               & C-G    & $0.667$   & $2.0$  & $0.5$  & $1.0$    \\
                                                  & 1D chain          &  $-$  & $-$   & $-$   & $-$   \\
                                                  & 2D S-L & $0.4035$   & $2.0$  & $0.165$  & $2.75$  \\ \hline
\end{tabular}\label{tab:table1}
\end{table}

When we compare the effects of the independence behavior and the anticonformity behavior on the critical point of the system, we find that the values of critical point $p_c$ with independent agents are smaller than those with anticonformist agent, both in the model on the complete graph and on the 2D square lattice. This condition means the 'noise effect' of the independent agent is more significant than the anticonformist agent. Therefore, the probability of undergoing a continuous phase transition is higher for the model with an independent agent. In other words, the systems with independent agents tend to experience a stalemate situation more than those with anticonformist agents. This result (the effect of independence behavior compared to anticonformity behavior on the critical point of the system is robust) is consistent to our previous results although the independent agent was involved in another opinion dynamics model under two different scenarios \cite{muslim2022opinion}. Our results suggest that existence of independence behavior in a society makes it more challenging to achieve consensus compared to the same society with anticonformists. Therefore, this paper corroborates our previous results that strong contrary behavior in various socio-political phenomena impacts a more stable state than disorganized free movement behavior.

\section*{CRediT authorship contribution statement}
\textbf{R. Muslim:} Conceptualization, Methodology, Software, Formal analysis, Investigation, Data Curation, Writing - original draft, Visualization. 
\textbf{S. A. Wella:} Supervision, Software, Formal analysis, Validation, Writing - review \& editing. \textbf{A. R. T. Nugraha:} Funding acquisition, Writing - review \& editing.

\section*{Declaration of Interests}
The authors declare that they have no known competing financial interests or personal relationships that could have appeared to influence the work reported in this paper.

\section*{Data availability}
The raw/processed data required to reproduce these findings are available to download from \href{https://github.com/muslimroni/majority-rule}{\textcolor{blue}{github}}.

\section*{Acknowledgments}
We thank BRIN Research Center for Quantum Physics for providing access to its mini-HPC to perform numerical simulations.

\bibliographystyle{elsarticle-num}
\bibliography{cas-refs}

\end{document}